\documentclass{article}
\usepackage{spconf,amsmath,graphicx}
\usepackage{booktabs}
\usepackage{algorithm,algorithmicx,algpseudocode}
\usepackage{adjustbox}
\usepackage{subfig}
\usepackage{makecell, tabularx, multirow}
\usepackage{diagbox}
\usepackage{array}
\usepackage{hyperref}
\usepackage{xcolor}
\usepackage{colortbl}
\definecolor{Ocean}{RGB}{129,194,234}

\definecolor{lightgray}{rgb}{0.85, 0.85, 0.85}
\definecolor{gray}{rgb}{0.7, 0.7, 0.7}

\title{AdvSV: An Over-the-Air Adversarial Attack Dataset \\for Speaker Verification}

\name{
\begin{tabular}{c}
Li Wang$^{1}$ \qquad  Jiaqi Li$^{1}$ \qquad Yuhao Luo$^{1}$ \qquad Jiahao Zheng$^{1}$ \qquad Lei Wang$^{2}$ \qquad \\Hao Li \qquad Ke Xu$^2$ \qquad Chengfang Fang$^2$ \qquad Jie Shi$^2$ \qquad Zhizheng Wu$^{1}$
\end{tabular}
}

\address{
\begin{tabular}{c}
$^1$School of Data Science, Shenzhen Research Institute of Big Data, \\ The Chinese University of Hong Kong, Shenzhen (CUHK-Shenzhen), China \\ $^2$IEEE Members, Singapore
\end{tabular}
}

\begin{document}
\ninept
\maketitle
\begin{abstract}
It is known that deep neural networks are vulnerable to adversarial attacks. Although Automatic Speaker Verification (ASV) built on top of deep neural networks exhibits robust performance in controlled scenarios, many studies confirm that ASV is vulnerable to adversarial attacks. \textit{The lack of a standard dataset is a bottleneck for further research, especially reproducible research}. In this study, we developed \textit{\textbf{an open-source adversarial attack dataset for speaker verification research}}. As an initial step, we focused on the over-the-air attack. An over-the-air adversarial attack involves a perturbation generation algorithm, a loudspeaker, a microphone, and an acoustic environment. The variations in the recording configurations make it very challenging to reproduce previous research. The AdvSV dataset is constructed using the Voxceleb1 Verification test set as its foundation. This dataset employs representative ASV models subjected to adversarial attacks and records adversarial samples to simulate over-the-air attack settings. The scope of the dataset can be easily extended to include more types of adversarial attacks. The dataset will be released to the public under the \textbf{CC BY-SA 4.0}. In addition, we also provide a detection baseline for reproducible research.
\end{abstract}
\section{Introduction}
\label{sec:intro}

An Automatic Speaker Verification (ASV) system is to decide whether a claimed identity is an impostor or a genuine speaker by comparing a presented utterance against an enrolled voice. Although deep learning has significantly enhanced the performance of ASV~\cite{jung2022pushing}, it is known that ASV is vulnerable to impersonation, replay, voice conversion, and speech synthesis as discussed in~\cite{wu2015asvspoof}. It is also known that deep neural networks are vulnerable to adversarial attacks \cite{survey_adversarial}. An \textit{\textbf{adversarial attack}} is an attack on deep learning by adding a perturbation. ASV systems built on top of deep neural networks are no exception.

There are two types of adversarial attacks in the context of ASV, namely digital  and over-the-air (OTA) adversarial attacks. A digital attack is to send a digital copy of an adversarial sample directly to an ASV system, and an OTA adversarial attack is to play a pre-generated adversarial sample in front of an ASV system. This process involves a loudspeaker, a microphone and acoustic environment or conditions (e.g. room reverberation and background noise). To perform attacks, the projected gradient descent (PGD)~\cite{madry2019deep} algorithm is commonly used to generate perturbations, and it was initially used in image classification. It has since been modified to attack ASV systems digitally. For instance, FoolHD~\cite{9413760} employs a multi-objective loss function to generate adversarial samples that are hard for humans to perceive. FakeBob~\cite{chen2019real} introduces a threshold estimation algorithm, combined with gradient estimation, to achieve black-box attacks. Zuo et al.~\cite{zuo22b_interspeech} propose a speaker-specific utterance ensemble method to enhance the generalization of adversarial attack samples.

Several studies have focused on OTA adversarial attacks.Xie et al.~\cite{xie2020realtime} use room impulse response to simulate room reverberation, increasing success rates of OTA adversarial attacks, employing the VCTK corpus. O'Reilly et al.\cite{oreilly2022effective} transform bonafide samples into adversarial ones with adaptive filtering using the VoxCeleb2 dataset. With a combination of Common Voice, CommanderSong~\cite{yuan2018commandersong} and LibriSpeech datasets, Zheng et al.~\cite{zheng2021blackbox} treats decision-only black-box adversarial attacks as a discontinuous large-scale global optimization problem, adaptively decomposing it into subproblems and collaboratively optimizing each one to find a solution. Both digital and OTA adversarial studies confirm security concerns. However, each study develops their dataset with a specific setting.

Various studies have addressed ASV system security concerns with countermeasures against adversarial attacks~\cite{wu2022adversarial,wu2022improving,wu2021voting,wu2023defending,peng2021pairing}. Adversarial-aware training approaches have been proposed in~\cite{wang2019adversarial,wu2020defense,pal2021adversarial} to enhance the ASV model's resilience to attacks. An x-vector-based attack signature has been proposed in~\cite{villalba2021representation} to detect adversarial perturbations. A diffusion-based approach has been proposed in~\cite{wu2023defending} to remove perturbations for discrimination. In these studies, the VoxCeleb1, VoxCeleb2, Speech Commands, TIMIT, ASVspoof2019, and Librispeech datasets are used. \textit{\textbf{Each individual study constructs its dataset for countermeasure research}}. Without a benchmark dataset, it is not feasible to perform benchmark comparisons, making reproducible research even more challenging.

Although there is a growing concern about the threat of adversarial attacks, \textit{the lack of a \textbf{benchmark dataset} is a bottleneck for further research, especially reproducible research}. Existing studies on adversarial attacks and countermeasures develop their own datasets, usually for specific purposes. To promote reproducible research, this work presents \textbf{an open-source dataset on adversarial attacks for speaker verification (AdvSV)}. OTA adversarial attacks involve a loudspeaker, a microphone, and acoustic environment or conditions (e.g. room reverberation and background noise). Due to the potential variations of OTA configurations, it is more challenging to conduct reproducible research. Hence, the AdvSV dataset is designed to serve that purpose. \textit{The focus of this study is over-the-air adversarial attacks, and the scope of the dataset can be easily extended to include more types of adversarial attacks.} The AdvSV dataset will be released to the public under the CC-BY license\footnote{http://creativecommons.org/licenses/by-sa/4.0/}.

\section{Over-the-air Adversarial Attack Dataset}
\label{sec:data}

\begin{figure}[htbp]
  \centering
  \includegraphics[width=0.45\textwidth]{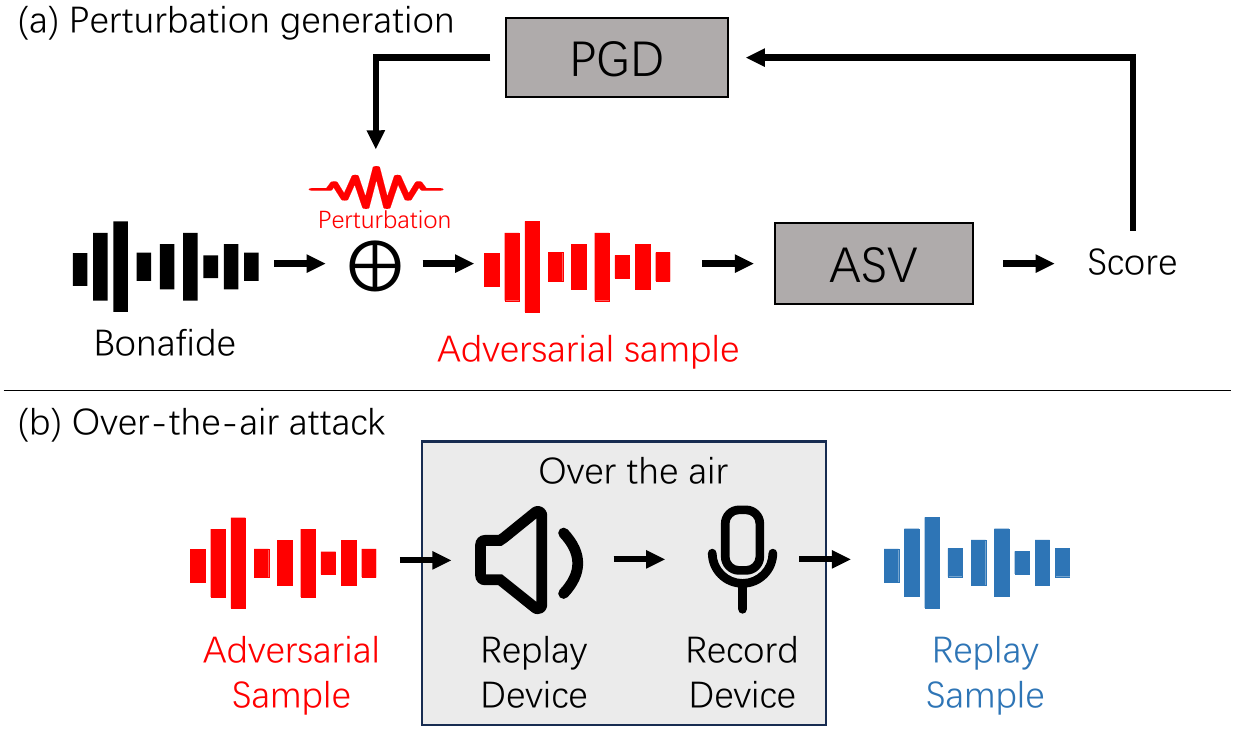}
  \caption{Illustration of an over-the-air adversarial attack, consisting of (a) perturbation generation and (b) over-the-air attack steps.}
  \label{fig:overtheair}
\end{figure}

This section presents a framework for producing the proposed AdvSV dataset. In this study, we focus on the over-the-air target attack, which modifies a sample to attack a specific target speaker's verification model. Fig.~\ref{fig:overtheair} presents an illustration of the over-the-air (OTA) adversarial attack, which consists of two steps: perturbation generation and OTA attack. Both steps will be described in this section.

\vspace{-1em}

\vspace{-2em}
\subsection{Perturbation generation}
\label{sec:pgd}

To synthesize adversarial samples, this study employs the project gradient descent (PGD)~\cite{madry2019deep} algorithm. The PGD algorithm is to add a perturbation to a testing sample, aiming to manipulate the ASV decision. The adversarial sample synthesis process is formulated in Eq.~\ref{eql:pgd}, 
\begin{equation}
\label{eql:pgd}
x_{t+1}=\prod_{x+S}\left(x_t+\alpha \cdot \operatorname{sign}\left(\nabla_x J\left(x_t, x_{enroll}, y\right)\right)\right.,
\end{equation}
where $x_{enroll}$ is an enrollment sample, $x_t$ is an adversarial sample at the t-th iteration, $y$ stands for a label (i.e. genuine or impostor), $\alpha$ is the step size, $S$ is the number of steps, $J$ signifies the loss function, and $sign$ is the sign function. When we apply the PGD algorithm, we assume the algorithm knows everything about the model including model architecture and parameters.

\begin{algorithm}[htbp]
  \caption{Ensemble PGD Attack}
  \label{alg:ensemble_pgd}
  \begin{algorithmic}[1]
    \Require{Victim models $\{M_1, M_2, \ldots, M_n\}$, enrollment sample $x_{enroll}$}
    \Ensure{Adversarial sample $\mathbf{x}_{\text{adv}}$}
    \Function{EnsemblePGD}{$\mathbf{x}$}
      \State Initialize $\mathbf{x}_{\text{adv}} \gets \mathbf{x}$
      \Repeat
        \For{$i \gets 1$ \textbf{to} $n$}
          \State Compute adversarial sample of \\ \qquad \qquad \qquad $M_i$: $x_{adv} \gets PGD(x_{enroll},x_{adv},y) $
        \EndFor
      \Until{Adversarial sample attacks all surrogate models}
      \State \Return $x_{adv}$
    \EndFunction
  \end{algorithmic}  
\end{algorithm}

The PGD algorithm presented in Eq.~\ref{eql:pgd} is to attack a single target ASV model. Different from the PGD algorithm, an ensemble PGD algorithm is to attack a few ASV systems at the same time~\cite{zhang2020blackbox}. The algorithm is presented in algorithm~\ref{alg:ensemble_pgd}. The idea of the ensemble PGD algorithm is to iteratively attack each victim model until the adversarial sample can spoof all the victim ASV models~\cite{zhang2020blackbox}. 

Both PGD and ensemble PGD algorithms require access to the victim model's parameters for gradient calculations. In practice, obtaining all the information about the target ASV model is \textit{not} feasible. A practical way to perform a \textit{transfer attack} is to synthesize an adversarial sample using one or a few known victim models and then use that adversarial sample to attack the target ASV system.

\textbf{Implementation details}: The \textit{PGD Attack} is configured with a step size ($\alpha$) of 0.004, 20 steps ($S$), and uses cosine similarity as the loss function. For the \textit{Ensemble PGD Attack}, three ASV models are used as victim models, while the remaining one serves as a test for transfer attacks. 

\vspace{-1em}
\subsection{Over-the-air attack setup}
\label{sec:replay_device}

An OTA adversarial attack involves a perturbation generation algorithm, a loudspeaker, a microphone, and a replaying environment. In this work, we simulated the OTA adversarial attack in a soundproof studio to reduce the impact of environmental noise and focus the dataset on the impact of perturbation generation, loudspeakers, and microphones. These three variables already result in a significant number of combinations.

We chose three types of loudspeakers and three types of recording devices (i.e., microphones). The high-end, medium-end, and low-end loudspeakers are priced at around \$300 USD, \$90 USD, and \$50 USD, respectively. For the recording devices, we chose mobile devices, which are common in our daily lives. The iOS, Android-high, and Android-low devices are priced at around \$900 USD, \$750 USD, and \$310 USD, respectively.

The distance and angle between the microphone and loudspeaker are other factors. In this study, we simplified this factor. The distance between the loudspeaker and microphone is set to 0.3 meters, and the angle is set to 90 degrees.

\vspace{-1em}
\subsection{Dataset}
\label{sec:sample}

To align with existing ASV research, we design the AdvSV dataset based on the \textit{Voxceleb1\footnote{https://www.robots.ox.ac.uk/~vgg/data/voxceleb/vox1.html} dataset}. The Voxceleb1 dataset is one of the most commonly-used dataset for speaker verification. We choose the Voxceleb1 verification set as the base set to generate adversarial samples. This set comprises 18,860 samples labeled with different speakers\footnote{Voxceleb1 verification test list: https://www.robots.ox.ac.uk/~vgg/data/\\voxceleb/meta/veri\_test.txt}. 

To reduce the burden of replay and recording, 25\% of the samples were retained (with the same speaker distribution). The proposed AdvSV dataset consists of a total of 314,496 samples\footnote{314,496 = 4368 x 4 victim models(section \ref{sec:exp_set}) x 2 attack methods x 3 replay devices x 3 recording devices}. Audio demo available on the webpage~\footnote{AdvSV dataset demo: https://advsv.github.io/}.


\section{Detection of Adversarial Attacks}
\label{sec:baseline}

\begin{figure}[htbp]
  \centering
  \includegraphics[width=0.49\textwidth]{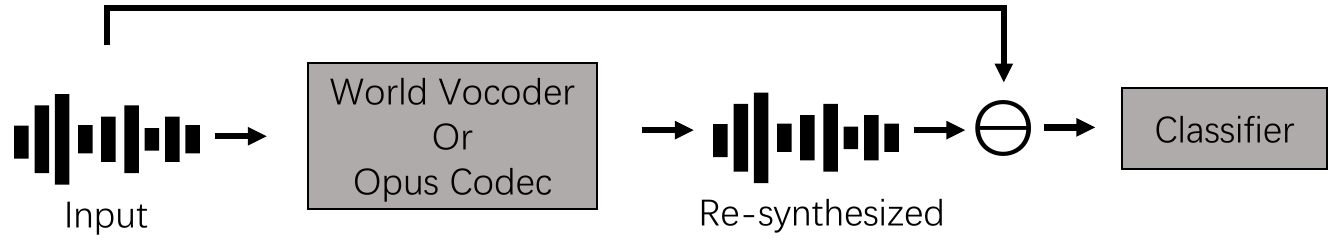}
  \caption{Framework of the baseline system to detect  adversarial attacks. `-' means subtracting re-synthesized spectrogram from the original spectrogram.}
  \label{fig:detection}
  \vspace{-2em}
\end{figure}

In this paper, we provide a countermeasure baseline based on the one-class classification method~\cite{wang2021dkucmri}, which is currently the mainstream approach for audio spoofing detection. Its pipeline is shown in Fig.~\ref{fig:detection}. During the training phase, we use the World~\footnote{https://github.com/JeremyCCHsu/Python-Wrapper-for-World-Vocoder} vocoder and Opus Codec~\footnote{https://opus-codec.org/} to re-synthesize the input audio, and subtract the re-synthesized audio from the input audio to remove irrelevant information for spoofing detection, such as speaker and speech content information. The subtracted audio is then fed into a one-class classifier. The world vocoder is used during the inference phase.

\section{Experiment}
\label{sec:exp}

\vspace{-1em}
\begin{table}[htbp]
  \centering
  \caption{Performance of ASV systems used as victim models in study. The performance is measured with Equal Error Rates (EER). }
  \adjustbox{max width=\columnwidth}{
    \begin{tabular}{llcc}
    \toprule
    Abbr. & Model & EER (Vox1 Full) & EER (Vox1 25\%) \\
    \midrule
    ECAPA&ECAPATDNN\cite{desplanques2020ecapatdnn} & 1.26\% & 1.26\% \\
    RawNet&RawNet3\cite{jung2022pushing} & 1.06\% & 1.17\% \\
    ResNet&ResNetSE34V2\cite{kwon2021ins} & 1.03\% & 0.92\% \\
    XVec&XVector\cite{snyder2018xvectors} & 2.08\% & 2.04\% \\
    \bottomrule
    \end{tabular}%
  \label{tab:victimmodels}%
  }
\end{table}%

\vspace{-1em}

\begin{table}[htbp]
  \centering
  \caption{Attack Success Rates (\%) of Various Attacks. S represents the surrogate model, and V represents the victim model.}
  \adjustbox{max width=\columnwidth}{
    \begin{tabular}{cccccc}
      \toprule
      Attack & \diagbox{S}{V} & RawNet & ECAPA & ResNet & XVec \\
      \midrule
      \multirow{4}{*}{PGD} 
      & RawNet & 98.9 & 10.6 & 8.8 & 14.1 \\
      & ECAPA & 54.9 & 100 & 43.5 & 62.5 \\
      & ResNet & 25.1 & 38.0 & 100 & 46.7 \\
      & XVec & 34.2 & 49.1 & 38.4 & 100 \\
      \midrule
      \midrule
      \multirow{4}{*}{\shortstack{Ensemble  \\ PGD}} 
      & w/o RawNet  & 86.0 & 100 & 100 & 100 \\
      & w/o ECAPA  & 100 & 62.3 & 100 & 100 \\
      & w/o ResNet & 100 & 100 & 62.5 & 100 \\
      & w/o XVec & 100 & 100 & 100 & 79.5 \\
      \bottomrule
    \end{tabular}
  }
  \label{tab:pgdensemble}
\end{table}

\begin{table*}[!ht]
  \centering
  \caption {Success rate(\%) of over-the-air adversarial attacks. Light gray areas represent PGD white-box attacks, and dark gray areas represent ensemble transfer attacks (i.e. blackbox attacks).}
  \adjustbox{max width=\textwidth}{
    \begin{tabular}{c|c|c|ccc|ccc|ccc|ccc|}
      \toprule
        & \multirow{3}{*}{Surrogate} & Victim & \multicolumn{3}{c|}{RawNet} & \multicolumn{3}{c|}{ECAPA} & \multicolumn{3}{c|}{ResNet} & \multicolumn{3}{c|}{XVec} \\
      \cline{3-15}
    Attack & & \multirow{2}{*}{\diagbox{\textcolor{purple}{Speaker}}{\textcolor{blue}{Mic}}} & \multirow{2}{*}{\textcolor{blue}{iOS}} & \textcolor{blue}{Android} & \textcolor{blue}{Android} & \multirow{2}{*}{\textcolor{blue}{iOS}} & \textcolor{blue}{Android} & \textcolor{blue}{Android} & \multirow{2}{*}{\textcolor{blue}{iOS}} & \textcolor{blue}{Android} & \textcolor{blue}{Android} & \multirow{2}{*}{\textcolor{blue}{iOS}} & \textcolor{blue}{Android} & \textcolor{blue}{Android} \\
      Method  &                            &    & &  \textcolor{blue}{High} &  \textcolor{blue}{Low} & & \textcolor{blue}{High} &  \textcolor{blue}{Low} & & \textcolor{blue}{High} &  \textcolor{blue}{Low} & & \textcolor{blue}{High} &  \textcolor{blue}{Low} \\
      \midrule
      \multirow{12}{*}{\rotatebox[origin=c]{90}{PGD}} 
      & \multirow{3}{*}{RawNet} 
      & \textcolor{purple}{High}     & \cellcolor{lightgray}31.9 & \cellcolor{lightgray}18.5 & \cellcolor{lightgray}26.6 & 9.2 & 8.5 & 9.0 &  8.3 & 8.1 & 8.0 &  12.2 & 11.8 & 11.3 \\
      & & \textcolor{purple}{Medium} & \cellcolor{lightgray}27.1 & \cellcolor{lightgray}18.0 & \cellcolor{lightgray}23.1 & 9.5 & 9.2 & 8.4 &  8.3 & 8.0 & 7.9 &  12.2 & 11.8 & 10.1 \\
      & & \textcolor{purple}{Low}    & \cellcolor{lightgray}20.6 & \cellcolor{lightgray}17.2 & \cellcolor{lightgray}20.5 & 7.9 & 8.2 & 7.5 & 7.5 & 7.4 & 8.1 & 10.3 & 10.8 & 7.4 \\
      \cline{2-15}
      & \multirow{3}{*}{ECAPA} 
      & \textcolor{purple}{High}     & 52.6 & 39.8 & 46.6 & \cellcolor{lightgray}100 & \cellcolor{lightgray}100 & \cellcolor{lightgray}100 &  41.3 & 40.5 & 39.0 &  56.9 & 53.3 & 47.6 \\
      & & \textcolor{purple}{Medium} & 49.8 & 39.7 & 44.0 & \cellcolor{lightgray}100 & \cellcolor{lightgray}100 & \cellcolor{lightgray}99.9 & 41.0 & 41.2 & 38.0 & 56.7 & 55.0 & 44.5 \\
      & & \textcolor{purple}{Low}    & 41.0 & 39.2 & 35.9 & \cellcolor{lightgray}100 & \cellcolor{lightgray}100 & \cellcolor{lightgray}99.7 & 36.8 & 39.3 & 32.7 & 49.6 & 54.2 & 36.6 \\
      \cline{2-15}
      & \multirow{3}{*}{ResNet} 
      & \textcolor{purple}{High}     & 24.8 & 19.3 & 21.5 & 31.3 & 29.5 & 30.4 & \cellcolor{lightgray}100  & \cellcolor{lightgray}99.8 & \cellcolor{lightgray}99.7 &  41.8 & 39.0 & 37.1 \\
      & & \textcolor{purple}{Medium} & 24.1 & 17.2 & 21.2 & 30.6 & 29.8 & 29.3 & \cellcolor{lightgray}99.9 & \cellcolor{lightgray}99.7 & \cellcolor{lightgray}99.7 & 41.1 & 38.4 & 33.3 \\
      & & \textcolor{purple}{Low}    & 18.6 & 17.2 & 16.8 & 26.0 & 28.0 & 24.7 & \cellcolor{lightgray}99.8 & \cellcolor{lightgray}99.9 & \cellcolor{lightgray}99.6 & 34.4 & 38.8 & 27.8 \\
      \cline{2-15}
      & \multirow{3}{*}{XVec} 
      & \textcolor{purple}{High}     & 35.3 & 25.1 & 30.3 & 41.3 & 39.9 & 40.2 & 36.7 & 35.3 & 34.2 & \cellcolor{lightgray}100 & \cellcolor{lightgray}100 & \cellcolor{lightgray}100 \\
      & & \textcolor{purple}{Medium} & 33.0 & 25.4 & 29.3 & 40.9 & 40.0 & 37.5 & 36.4 & 35.1 & 32.4 & \cellcolor{lightgray}100 & \cellcolor{lightgray}100 & \cellcolor{lightgray}99.7 \\
      & & \textcolor{purple}{Low}    & 27.8 & 23.9 & 23.5 & 36.3 & 39.1 & 33.5 & 30.9 & 34.3 & 27.5 & \cellcolor{lightgray}100 & \cellcolor{lightgray}100 & \cellcolor{lightgray}98.7 \\
      \midrule
      \midrule
      \multirow{12}{*}{\rotatebox[origin=c]{90}{Ensemble PGD}} 
      & \multirow{3}{*}{w/o RawNet} 
      & \textcolor{purple}{High}     & \cellcolor{gray}84.1 & \cellcolor{gray}70.2 & \cellcolor{gray}77.9 & 100 & 100 & 99.9 & 99.4 & 97.7 & 98.0 & 99.3 & 99.0 & 98.2 \\
      & & \textcolor{purple}{Medium} & \cellcolor{gray}81.3 & \cellcolor{gray}71.1 & \cellcolor{gray}77.2 & 100 & 100 & 100 & 99.4 & 99.0 & 98.0 & 99.5 & 99.5 & 96.7 \\
      & & \textcolor{purple}{Low}    & \cellcolor{gray}78.8 & \cellcolor{gray}67.9 & \cellcolor{gray}70.5 & 100 & 100 & 100 & 98.6 & 98.4 & 94.8 & 98.7 & 98.9 & 93.8 \\
      \cline{2-15}
      & \multirow{3}{*}{w/o ECAPA} 
      & \textcolor{purple}{High}     & 77.4 & 65.3 & 69.7 & \cellcolor{gray}53.8 & \cellcolor{gray}52.9 & \cellcolor{gray}53.4 & 99.8 & 98.7 & 98.6 & 98.7 & 97.3 & 95.1 \\
      & & \textcolor{purple}{Medium} & 74.9 & 57.3 & 68.8 & \cellcolor{gray}54.3 & \cellcolor{gray}54.0 & \cellcolor{gray}51.5 & 99.7 & 99.4 & 98.9 & 98.9 & 98.5 & 91.8 \\
      & & \textcolor{purple}{Low}    & 69.1 & 55.9 & 63.6 & \cellcolor{gray}50.5 & \cellcolor{gray}49.9 & \cellcolor{gray}45.8 & 99 & 98.8 & 95.5 & 95.9 & 96.7 & 85.5 \\
      \cline{2-15}
      & \multirow{3}{*}{w/o ResNet} 
      & \textcolor{purple}{High}     & 91.6 & 75.4 & 85.7 & 100 & 99.9  & 99.8 & \cellcolor{gray}58.1 & \cellcolor{gray}56.4 &  \cellcolor{gray}55.4 & 98.8 & 98.0 & 95.6 \\
      & & \textcolor{purple}{Medium} & 87.0 & 76.5 & 84.4 & 99.9 & 99.9 & 99.9 & \cellcolor{gray}57.8 & \cellcolor{gray}58.2 &  \cellcolor{gray}54.2 & 99.0 & 98.9 & 94.3 \\
      & & \textcolor{purple}{Low}    & 86.9 & 73.3 & 80.6 & 99.9 & 99.7 & 99.4 & \cellcolor{gray}54.9 & \cellcolor{gray}54.9 &  \cellcolor{gray}49.7 & 97.6 & 97.9 & 89.0 \\
      \cline{2-15}
      & \multirow{3}{*}{w/o XVec} 
      & \textcolor{purple}{High}     & 89.1 & 76.4 & 82.6 & 100  & 99.9 & 99.9 & 91.6 & 85.6 & 85.9 & \cellcolor{gray}71.4 & \cellcolor{gray}68.5 & \cellcolor{gray}63.6 \\
      & & \textcolor{purple}{Medium} & 83.5 & 72.7 & 81.8 & 99.9 & 99.9 & 99.8 & 91.6 & 89.8 & 87.5 & \cellcolor{gray}71.9 & \cellcolor{gray}71.4 & \cellcolor{gray}62.2 \\
      & & \textcolor{purple}{Low}    & 84.1 & 68.8 & 77.9 & 99.8 & 99.9 & 99.4 & 87.6 & 85.4 & 78.1 & \cellcolor{gray}66.9 & \cellcolor{gray}67.4 & \cellcolor{gray}55.5 \\
      \bottomrule
    \end{tabular}
  }
  \label{tab:replaypgdensemble}
\end{table*}

\begin{table*}[!ht]
  \centering
  \caption{Detection results of over-the-air adversarial attacks (EER\% $\downarrow$). The results for the four victim models are averaged to obtain the results for both PGD and ensemble PGD attacks. The training set for row 4a and 4b are bonafide samples go through the OTA process, while training samples of other rows use bonafide samples. The setting of testing set is indicated in 2nd column. The column `overall' pools all the testing samples together to calculate EERs.}
  \adjustbox{max width=\textwidth}{
    \small
    \begin{tabular}{c|c|c|ccc|ccc|ccc|c}
      \toprule
      \multirow{1.5}{*}{Row} & \multirow{1.5}{*}{Spoof Sample of} & \textcolor{purple}{Speaker} & \multicolumn{3}{c|}{\textcolor{purple}{High}} & \multicolumn{3}{c|}{\textcolor{purple}{Medium}} & \multicolumn{3}{c|}{\textcolor{purple}{Low}} & \multirow{3}{*}{Overall} \\
      \cline{3-12}
       \multirow{1.5}{*}{Index} & \multirow{1.5}{*}{the Test Set} & \multirow{2}{*}{\diagbox{Attack}{\textcolor{blue}{Mic}}} & \multirow{2}{*}{\textcolor{blue}{iOS}} & \textcolor{blue}{Android} & \textcolor{blue}{Android} &  \multirow{2}{*}{\textcolor{blue}{iOS}} & \textcolor{blue}{Android} & \textcolor{blue}{Android} & \multirow{2}{*}{\textcolor{blue}{iOS}} & \textcolor{blue}{Android} & \textcolor{blue}{Android} &\\
       & &  &                      & \textcolor{blue}{High}    & \textcolor{blue}{Low}     &                       & \textcolor{blue}{High}    & \textcolor{blue}{Low}     &                    & \textcolor{blue}{High}    & \textcolor{blue}{Low}  &   \\
      \midrule
      1 & OTA   &  NA & 4.90 & 4.97  & 4.63 & 8.24 & 10.04 &6.53 &4.68  &7.05 & 4.73   &6.66     \\ 
      \midrule
      \midrule
      2a & \multirow{2}{*}{Digital} & PGD &  &	\multirow{2}{*}{N/A} &	 &	 &	\multirow{2}{*}{N/A} &	 	& &	\multirow{2}{*}{N/A} & &	66.73  \\
      \cline{3-3}\cline{13-13}
      2b & & Ensemble &  &	 &	 &	 &	 &	 &	 	&  &	 &	69.83  \\
      \midrule
      \midrule
      
      3a & \multirow{2}{*}{OTA} & PGD & 3.66 &	1.24 &	1.14 &	1.27 &	1.02 &	0.67 	&4.27 &	3.85 &	4.06 &	3.49  \\
      \cline{3-13}
      3b & & Ensemble & 4.65 &	1.09 &	1.13 &	1.58 &	1.07 &	0.48 &	4.06 	& 2.60 &	3.20 &	3.20  \\

      \midrule
      \midrule
      4a &   OTA   & PGD & 55.38 &	19.96 &	9.13 &	4.74 &	2.27 &	3.74 &	65.13 &	34.48 &	47.67 &	33.53  \\
      \cline{3-13}
      4b & (Bonafide: OTA)& Ensemble & 61.39 &	15.32 &	8.34 &	6.76 &	2.39 &	1.71 &	66.76 &	24.54 &	44.81 &	34.61  \\
      \bottomrule
    \end{tabular}
  }
  \label{tab:baseline_adv_replay}
\end{table*}

\subsection{Experimental Setup}
\label{sec:exp_set}
\textbf{ASV Models:} We use four state-of-the-art ASV systems to generate adversarial samples. The ASV models are trained and tested with open-source toolkit\footnote{\href{https://github.com/clovaai/voxceleb\_trainer}{https://github.com/clovaai/voxceleb\_trainer}}. The ASV systems are trained on Voxceleb2 and tested on either the full dataset or a 25\% subset of Voxceleb1. The EERs are summarized in Table~\ref{tab:victimmodels}. It is observed that the difference between EERs on the full dataset and the subset is trivial. We presume the subset can represent the distribution of the full dataset, and hence we use the subset to construct the AdvSV dataset.

\textbf{Detection Model:} The model is trained on the bonafide or replayed bonafide samples from the Voxceleb2 dataset, and is tested on the AdvSV dataset, assuming the detector has \textit{zero} knowledge or \textit{limited} knowledge of the adversarial attacks.

\textbf{Evaluation metric:} We use attack success rates and equal error rates(EER)~\cite{KARNAN20111565} as evaluation metrics to evaluate adversarial attack performance and detection performance. The success rate is defined as, $\textit{Attack Success Rate} = \frac{\text{Number of Successful Attacks}}{\text{Number of Attacks}}$

\vspace{-1em}
\subsection{Digital Adversarial Attack}

We first examine the performance of digital adversarial attacks with the AdvSV dataset. The results are presented in Table~\ref{tab:pgdensemble}. It is observed that if a surrogate model is the same as the victim model (i.e. \textit{white-box attack}), the success rates are always high, 100\% or close to 100\%. \textit{Note that a white-box attack is expecting to achieve 100\% success rate if there are enough PGD steps. In the experiments, we keep the same PGD steps for a fair comparison.} Ensemble attacks achieve higher transferability than the single PGD attack. \textit{Transferability} measures the performance of transfer attacks. In particular, with the PGD algorithm, the success rates are 10.6\%, 8.8\% and 14.1\% when using RawNet to transfer attack ECAPA, ResNet and XVec, respectively. However, the ensemble PGD will increase success rates of transfer attacks to 62.3\%, 62.5\% and 79.5\% for ECAPA, ResNet and XVec, respectively.

In summary, the success rates of PGD white-box attack are as high as 100\% or close to 100\%. Ensemble attacks can dramatically increase the success rates of transfer attacks.

\vspace{-1em}
\subsection{Over-the-Air Adversarial Attack}

We then assess the performance of over-the-air(OTA) adversarial attacks with the AdvSV dataset. Table~\ref{tab:replaypgdensemble} presents the success rates of OTA adversarial attacks. 

For the PGD white-box attacks, the RawNet system behaves differently from other systems. The success rates are considerably lower than those of other systems. This is because we set the same fixed number of PGD steps for all systems, and RawNet requires more PGD steps to achieve a higher success rate. From the loudspeaker perspective, a high-end speaker produces higher success rates than a low-end speaker. Similarly, a high-end phone/microphone achieves higher success rates than low-end phones or microphones. Additionally, the success rates of transfer attacks are considerably lower than those of white-box attacks. However, the success rates of transfer attacks still fall within the range of 30\% - 60\%, with RawNet being an exception.

For the ensemble attacks, the success rates of transfer attacks are considerably higher than those of PGD transfer attacks. For example, when attacking the XVec system, the ensemble PGD can increase the success rates from the range of 7.4\% - 56.9\% to the range of 55.5\% - 71.9\%. The phenomenon in RawNet is similar to that in PGD, which is due to the fixed number of PGD steps. Similar to PGD attacks, high-end loudspeakers or phones give higher success rates than low-end ones.

In summary, when facing OTA adversarial attacks, ASV systems are still vulnerable to transfer attacks, even if the ensemble PGD algorithm has no access to the target system. Different OTA settings could result in different success rates.

\vspace{-1em}
\subsection{Detection of Over-the-Air Adversarial Attacks}
\label{sec:baseline}

Last but not least, we provide a baseline for detecting both digital and OTA adversarial attacks with the AdvSV dataset. The detection results are presented in Table~\ref{tab:baseline_adv_replay}.

In row 1, the classifier is trained on bonafide data without the OTA process, but the testing data is processed with the OTA process. The overall detection EER is 6.66\%. Rows 2a and 2b present the detection of digital adversarial attack samples. In comparison with row 1, the EER increases to 66.73\% and 69.83\% for PGD and ensemble attacks, respectively. Rows 3a and 3b present the detection results of OTA adversarial attacks. Note that the training data are bonafide samples \textit{without} the OTA process. The overall EER is 3.49\% and 3.20\% for PGD and ensemble attacks, respectively. The only difference between row 3a/3b and row 1 is whether perturbations are added. \textbf{The results indicate that the OTA process plays a more important role in the detection of attacks}. Rows 4a and 4b use the same testing set as that in 3a and 3b, however, the training set of 4a and 4b \textit{are bonafide samples passing through the OTA process}. In comparison to 2a and 2b, both training and testing sets of 4a and 4b go through the OTA process. Both settings have considerably high EERs (i.e. higher than 30\%).

From the detection results, it suggests that \textbf{\textit{detecting the adversarial attacks or adversarial perturbations is a more challenging task} than detecting whether an audio sample has gone through the OTA process}.

\section{Conclusions and Future Work}

We designed an over-the-air adversarial attack dataset for speaker verification, called the \textit{AdvSV dataset}, which will be released under the CC BY-SA 4.0 license. To develop the dataset, we used three loudspeakers, three microphones, two perturbation generation algorithms and four state-of-the-art ASV systems. \textit{\textbf{In terms of adversarial attack success rate, the dataset presents a genuine problem; the success rates can be higher than 50\% in transfer blackbox attacks}}. In terms of baseline detection performance, there is still a long way to go to develop a successful countermeasure. In future work, we will continue to expand the dataset by considering more realistic product scenarios.

\vfill\pagebreak

\bibliographystyle{IEEEbib}
\bibliography{strings,refs}

\end{document}